\documentclass[conference]{IEEEtran}

\usepackage{graphicx,latexsym,epsfig,amssymb,amsmath,enumitem}
\usepackage{algorithmic}
\usepackage{verbatim}
\usepackage[ruled]{algorithm}
\usepackage{caption}
\usepackage{multirow}
\usepackage{subfig}
\usepackage{float}

\begin{document}

\title{ Geographic Trough Filling for
Internet Datacenters}  

\author{
\authorblockN{\small Dan Xu and Xin Liu}
\authorblockA{ Computer Science Department,
 University of California, Davis,
 \{danxu, xinliu\}@ucdavis.edu} }

%

\maketitle

\begin{abstract}
To reduce datacenter energy consumption and cost, current practice has considered  demand-proportional resource provisioning schemes, where
servers are turned on/off according to the load of requests.
 Most existing work considers instantaneous (Internet) requests only, which are explicitly or implicitly assumed to be delay-sensitive. On the other hand, in datacenters, there exist a vast amount of delay-tolerant jobs, such as background/maintainance jobs.
In this paper, we explicitly differentiate delay-sensitive jobs and delay tolerant jobs. We focus on the problem of using
delay-tolerant jobs to fill the extra capacity of datacenters, referred to as trough/valley filling.  Giving a higher priority
to delay-sensitive jobs, our schemes complement to most existing
demand-proportional resource provisioning schemes. Our goal is to design intelligent trough filling mechanisms that are  energy efficient and also achieve good delay performance.
Specifically, we propose  two  joint  dynamic speed scaling  and traffic shifting schemes, one subgradient-based and the other queue-based.  Our schemes assume little
statistical information of the system, which is usually difficult to obtain
in practice.
In both schemes, energy cost saving comes from dynamic speed scaling, statistical multiplexing, electricity price diversity, and service efficiency diversity. In addition, good delay performance is achieved in the queue-based scheme via load shifting and capacity allocation based on queue conditions.  Practical issues that may arise in datacenter networks   are considered, including capacity and bandwidth constraint, service agility constraint, and load shifting cost. We use both   artificial and real datacenter traces  to evaluate the proposed schemes.

 \end{abstract}


%
\maketitle

\section{Introduction}

The fast proliferation of  cloud computing
has promoted rapid growth of large-scale commercial
 datacenters.   Major service providers often deploy tens
 to hundreds of datacenters distributed nationwide or  even worldwide, referred to
 as Internet-scale datacenters (IDC). Because electricity bill contributes to a large portion of IDC operational expenditure,  there have been lots of efforts towards reducing IDC energy consumption/cost.

Researchers have  considered designing `load-aware' IDCs, e.g., in  \cite{yiyuchen}\cite{gongchen}\cite{caltech1}.
 The key idea is to
provision servers according to
the load of Internet requests. Extra servers are shut down or scheduled in sleeping mode to save energy.
In this paradigm, a major challenge is   to properly size an IDC, i.e., to determine the number of active servers, and in the meantime guarantee the service requirement. For example, in \cite{gongchen}, the authors propose to predict the load of windows live messengers and provision servers accordingly. In \cite{caltech1}, the authors estimate the current load,  and design online server provisioning schemes to reduce energy and server state transition cost, which is referred to as dynamic ``right sizing''.

In the above-mentioned work, service requests are typically delay-sensitive, i.e., requiring a short delay and low drop rate. Such applications include searching or signing in a messenger.
   When the load is lower, more servers would be turned off to save energy.
   However, in practice, an IDC operator may be reluctant to turn off  servers in a large scale even at a low load of requests.
   One reason is that turning on/off servers frequently  affects QoS and long term system reliability,  as considered in \cite{yiyuchen}. But the foremost reason is that
   there are also a large amount of background or
  maintenance jobs in  IDCs to process, e.g.,
    searching engine tunes ranking algorithms. Thus, the ``extra'' capacity can be utilized to process the background analytical jobs. This is referred to as \emph{trough/valley filling}.

Trough filling has not been studied thoroughly.
In this paper, we focus on intelligent trough filling.  We assume a given capacity provisioning and  scheduling mechanism for delay-sensitive jobs (DSJs), e.g., those proposed in \cite{yiyuchen}\cite{gongchen}\cite{caltech1}\cite{aotang2}\cite{raolei}\cite{Narayan}. We decide how to use load shifting and dynamic speed scaling to control  delay tolerant jobs (DTJs), e.g., background analytical jobs.
 On one hand,
DTJ load is high and thus its energy cost is considerable. On the other hand, it is  desirable to assure a good delay performance for DTJs.  The goal of intelligent trough filling is thus to achieve energy efficiency as well as good delay performance (or at least guarantee the queue stability) for DTJs.

Intelligent trough filling needs to accommodate the following issues.  First, the overall capacity of a datacenter is  likely to be random, e.g., due to server failure. Second, capacity demand of DSJs, such as Internet requests, varies   due to dynamic  load.
Given the higher priority of DSJs,  available capacity  for DTJs is random and hard to predict or learn in statistics. Meanwhile,  the demand of DTJs is also likely to be dynamic.

Further, in order to  consider a set of geographically distributed IDCs,
there are additional constraints.
First, load shifting is constrained by the bandwidth available between IDCs. In our setting, similar to capacity, bandwidth is prioritized for shifting DSJs, and thus results in a random `residual bandwidth' for DTJs. Second,  electricity prices diversity and dynamics
 bring challenges as well as opportunities, e.g., in  price-aware load shifting~\cite{cutting}-\cite{Jain}, in the context of trough-filling.  Third, due to heterogenous service agility, different classes of DTJs may require different sets of IDCs. Moreover,  different IDCs maybe heterogenous in service rates and energy consumption for each type of DTJs.
We consider these issues and address the above challenges in this paper.

In this paper, our goal is to
 design intelligent trough filling mechanisms, that achieve both energy efficiency and good delay performance.
We design  joint dynamic speed scaling and load shifting schemes.  Specifically, we make the following contributions:
\begin{itemize}
\item We focus on   trough filling in distributed IDCs, which compliments the current work on load-aware capacity provisioning, or price-aware load shifting.

\item We consider practical  issues in IDCs, such as dynamic capacity and bandwidth constraints, dynamic demand, and heterogenous service agility and service rates.

\item We first propose a stochastic subgradient based trough filling scheme, named SSTF, with the objective of minimizing energy and shifting cost  while stabilizing the DTJ queues. The proposed algorithm does not need
  underlying probability of system states, which is usually difficult to estimate.

 \item
We further propose a queue-based trough filling algorithm, called QTF, which does not need any statistical system information. We show the QTF  achieves  desirable performance in terms of  cost and queue delay.

\item We   discuss on  how to incorporate capacity provisioning and QoS assurance for DSJs into our proposed SSTF and QTF.

\item We use both synthetic traffic trace and real datacenter traffic trace to evaluate our proposed schemes. Simulation results show that QTF outperforms SSTF significantly in both cost and queue delay.

\end{itemize}

 The rest of paper is organized as follows.
In Section~\ref{sec:related}, we  survey  related work. In
Section~\ref{sec:model}, we describe the system model. In
Section~\ref{sec:ss}, we present  stochastic subgradient based trough filling scheme.  We further propose a queue based trough filling scheme in
Section~\ref{sec:qba}. We also discuss how to extend the schemes to DSJs and implementation issues in Section~\ref{sec:joint}. We evaluate our proposed schemes in
Section~\ref{sec:evaluation}, and conclude in
Section~\ref{sec:con}.

\section{Related work}
\label{sec:related}

Industry and academic research community have paid much attention
to reducing datacenter energy consumption and cost.
Solutions are considered in all spectra,
 including power-efficient chip, cooling system,
deployment, and many others.

Our work  complements to load-aware server provisioning or
power-proportional
design~\cite{yiyuchen}-\cite{SOSP2001}. Such works focus on server or resource provisioning based on load of
Internet requests, with service level agreement~{SLA} or other QoS
metrics assured. For example, in \cite{yiyuchen}, the authors propose server
provisioning and dynamic speed/voltge scaling schemes for a data center, through load
prediction and feedback control. Load prediction-based
 server provisioning and load dispatch is proposed in
\cite{gongchen} for connection-intensive Microsoft datacenter.
Online resource or server provisioning schemes are designed in
\cite{spaa}\cite{caltech1}. In \cite{caltech1}, the authors consider
a relative large time interval such that current load of requests
can be estimated. Server state transition cost is also considered. Furthermore, the authors also consider the impact of trough filling on energy saving by the proposed scheme though simulations.
Queue based server provisioning and Lyaponuv optimization based
performance establishment is proposed in~\cite{neely}.
Although the  Lyaponuv optimization technique is also used to show performance of the queue-based scheme,
our problem is different, i.e., we consider trough-filling, with  cross-datacenter
 load shifting and capacity provisioning.
In \cite{SOSP2001}, the authors propose an
economic framework which maximizes the total profit of resource
provisioning for all requests.

Many other power management schemes  for a datacenter
have been proposed, e.g.,  in \cite{osdi94}-\cite{ISCA2007}.
Dynamic speed/voltge scaling   saves power consumption of a
processor by adjusting the frequency based on the instantaneous load
demand,
  e.g. in \cite{osdi94}-\cite{lijunchen}, which can also be considered
  as load-aware resource provisioning. However, most of the work
  only considers a single processor.  In \cite{aotang2}, the authors use MDP to find optimal stationary DVS and load balancing policy to reduce service cost. In this paper, we use DVS as a part of control mechanism for trough filling in IDCs.
Another popular scheme is  virtualization and server consolidation,
e.g., in \cite{IPDPS}-\cite{xiaoqiao2}, which can reduce
the traffic dynamics by consolidating applications, and thus reduce the number of
active servers.  There are also some other works on
datacenter-level power management, such as workload
decomposition~\cite{decomposition}, optimal power allocation for
servers with total power budget~\cite{SIGMETRICS09}, model predictive control (MPC) theory based  hierarchical power control~\cite{wangxiaorui},  and other
techniques~\cite{shenkai}\cite{ISCA2007}\cite{nano:09}.

Most recently, cross-IDC power and cost optimization that exploits
geographic diversity has received significant attention,  e.g., in
\cite{cutting}-\cite{Jain}. The key idea is to shift
requests to IDCs with lower electricity prices to reduce cost. The tradeoff is the extra delay caused by traffic shifting. Thus, in \cite{raolei}\cite{Narayan},
the authors consider response time as the constraint. In \cite{caltech2}\cite{Jain}, the
authors consider shifting cost as the revenue loss incurred by extra
delay. Our work can also leverage price diversity, i.e., by
filling cheap troughs of IDCs. The difference is that since
background jobs are delay tolerant, our capacity provisioning and
load shifting schemes also exploit the temporal price diversity,
in addition to geographic diversity. In a recent work~\cite{mneely}, the authors use energy storage systems to leverage the temporal price dynamics to cut the energy cost, but for a single datacenter.

 We refer readers to \cite{liu:ICDCSW:09} for a survey and \cite{james} for  discussions on
challenges and issues in IDC power management.

\section{System Models}
\label{sec:model}

\subsection{The IDC and server model   }
We consider one service provider with a set of $N$ IDCs in  different
 locations.
An IDC $i$ has $K_i^{max}$ homogenous servers. We consider a time slotted system, where the  slot length can be from hundreds of milliseconds to minutes. We assume in each slot $t$, the number of active
servers of an IDC $i$ is
 fixed and is denoted  by $K_i^t$. Note that
$K_i^t$ varies over time,  due to either dynamic service provisioning (e.g., those proposed in \cite{gongchen}\cite{caltech1}\cite{raolei})
 or server failure.

An active server operates  at a CPU speed of $s$. Following the models
in \cite{soda07}\cite{aotang}\cite{Stanojevic}, we normalize $s$, i.e.,
$0\leq s\leq 1$, where 0 represents the idle state of an active server,
and 1 represents the maximum frequency. We define the capacity of an
IDC $i$ as the sum of speed of all active servers. If each server
runs at the same speed $s$, the total capacity in time slot $t$ is
$K_i^ts$. Clearly, the maximum capacity with $K_i^t$ servers is
 $K_i^t$.  In this paper, we consider CPU resource as the the main
bottleneck and focus on CPU capacity scheduling. The impact of other
equipments, i.e.,
 memory and I/O, will  be considered in heterogenous service rates, as discussed in subsection~\ref{sericemodel}.
Because scaling up/down the speed $s$ of an
active server  only takes several microseconds~\cite{aotang}\cite{unsal}, which is negligible,
 dynamic speed scaling can be conducted instantaneously
 in each time slot.

\subsection{Workload  model}
We consider two categories
 of demand: delay sensitive jobs (DSJs), e.g., searching, email login in, or messenger sign up, and delay tolerant jobs (DTJs), e.g., background analytical jobs.
 DSJs enjoy a higher priority on capacity allocation.
 The remaining
 capacity can be utilized by the DTJs.
  Since
 the load of DSJs is usually  dynamic, capacity demand of DSJs in an IDC $i$ in each slot is considered random. We use
 $S_{i0}^t$ to denote the capacity allocated to DSJs at IDC $i$ in slot $t$. We assume  $S_{i0}^t$  is given, based on some existing
 load-aware capacity provisioning schemes. Available capacity  for DTJs in IDC $i$ is thus $K_i^t-S_{i0}^t$.

For DTJs, they can be further divided into different
classes to capture their different resource requirements.  We consider there are in total $M$ different classes of DTJs in  the $N$ IDCs.
If the   same kind of DSJs, e.g., tuning webpage ranking algorithms,  originates (first arrives) at  different IDCs, we treat them as different classes.
This is because they may have different sets of IDCs to be shifted to due to distance constraints.    For DTJ $j$, it first originates at an IDC $i$.
Let   $D_j^t$  denote the traffic or load size of DTJ $j$ in time slot $t$.
$D_j^t$ is a random variable.  We do not make
 assumptions on its distribution.

\subsection{Models for load shifting and service  }
\label{sericemodel}

Although a DTJ $j$ originates at an IDC $i$, we can shift
the traffic  to other IDCs, e.g., to exploit their available capacity or  lower
electricity prices.  Note that cross-IDC load shifting is practically feasible due to negligible shifting time delay~\cite{Stanojevic}, which has been widely considered,
e.g., in \cite{cutting}-\cite{jamesblog}.
Load shifting has practical constraints. First, due to limited service agility of IDCs, a class of DTJ $j$
can potentially be served by  only a subset of IDCs.  Let $\Gamma_j$ denote the set of IDCs that can serve DTJ $j$, which is different for different classes of DTJs.  DTJ $j$ can only be shifted to IDC $i^{'}$, where $i^{'}\in\Gamma_j$.
Second, bandwidth between IDCs is limited. Moreover,  due to potentially load shifting for DSJs,
which also requires a high priority of  bandwidth provisioning, available bandwidth
for DTJs is limited and dynamic.  This consideration is similar to that in a very recent work~\cite{Laoutaris}, where the authors develop a system to rescue unutilized network bandwidth  for shifting the non-real-time bulk data, e.g., backup data.
We use $B_{ii^{'}}^t$ to denote the
available bandwidth from IDC $i$ to $i^{'}$ for DTJs
in slot $t$.  $B_{ii^{'}}^t$ varies over time,  and can be set in an appropriate value to prevent significant network delay.  Note when two IDCs  have
 limited connections
or a long distance  such that load shifting is not desirable,
  $B_{ii^{'}}^t$ can be set as 0 for all time slots.
 Let $D_{jii^{'}}^t$ denote the traffic of DTJ $j$ shifted from IDC $i$ to $i^{'}$.
 Further let $\Upsilon_{ii^{'}}$ denote the set of DTJs that first
  arrive at IDC $i$ and can be served by IDC $i^{'}$. We have $\sum_{j\in \Upsilon_{ii^{'}}}D_{jii^{'}}^t\leq B_{ii^{'}}^t$
    as the load shifting constraint.

 For an IDC $i\in  \Gamma_j$, it allocates a certain
capacity to DTJ $j$ in time slot $t$, denoted by
  $S_{ij}^{t}$.  We have
$\textbf{S}^{t}=\{S_{ij}^t|j=1,\ldots, M,  i\in  \Gamma_j \}$, as the capacity allocation matrix, which is our control variable.
An IDC $i$ may serve multiple DTJs. Let $\Pi_i$ denote
     the set of all DTJs served by an IDC $i$. Obviously, we have the capacity allocation constraint as $\sum_{j\in \Pi_i}S_{ij}^t\leq {K}_{i}^{t}-S_{i0}^t$.

With capacity
     $S_{ij}^{t}$, DTJ $j$ receives a certain service rate.
      We use  the  $R_{ij}(S_{ij}^t)$ as the service rate function on the capacity.
       For simplicity,   we consider $R_{ij}()$  as a linear  function of $S_{ij}^t$, i.e.,  $R_{ij}(S_{ij}^t)=r_{ij}S_{ij}^{t}$.
      The unit service rate $r_{ij}$  is heterogenous for different pairs of DTJ $j$ and IDC $i$.  This is because different
    DTJs may require different memory, I/O resource, etc.
Load shifting and dynamic speed scaling are coupled. The amount of traffic of DTJ $j$ shifted from IDC $i$
 to $i^{'}$ depends on the capacity allocated at IDC $i^{'}$. Thus we have $D_{jii^{'}}^t\leq r_{i^{'}j}S_{i^{'}j}^t$. Since both energy and load shifting cost  increase with  $S_{i^{'}j}^t$,   we have $D_{jii^{'}}^t= r_{i^{'}j} S_{i^{'}j}^t$.

 The unfinished jobs of a DTJ $j$ are buffered in a queue at the IDC where DTJ $j$ originates. Let $Q_j(t)$ denote the queue  in time $t$, the queue dynamics of
   DTJ $j$ can be written as
\begin{equation}
Q_j(t+1)=\max\left[Q_j(t)- \sum_{i\in \Gamma_j}r_{ij}S_{ij}^t,
0\right]+   D_j^t, \label{powerserver1}
\end{equation}
where $\sum_{i\in \Gamma_j}r_{ij}S_{ij}^t$ is the total service rate a DTJ $j$
receives in slot $t$.

\begin{table}
\renewcommand{\arraystretch}{1}
\caption{ Main Notations } \label{table:notions}
\scriptsize
\begin{center}
\begin{tabular}{c|c}
\hline
\ $K_i^t$  & Number of active servers of IDC $i$ in slot $t$  ($K_i^{\omega}$ for state  ${\omega}$)\\
\ $D_j^t$ &  Traffic arrival  of DTJ $j$ in slot $t$   \\
\ $B_{ii^{'}}^t$ &  Bandwidth constraint for DTJs between IDC $i$ and $i^{'}$ in slot $t$  \\
\ $\Upsilon_{ii^{'}}$ & Set of different types of DTJs shifted from IDC $i$ to $i^{'}$ \\
\ $\Gamma_j$ & Set of IDCs that can serve DTJ $j$ \\
 \ $\Pi_i$  & Set of different types of DTJs served by IDC $i$ \\
  \ $S_{ij}^{t}$ & Capacity/speed allocated by IDC $i$ ($i\in \Gamma_j$) to DTJ $j$ in slot $t$ \\
 \   $S_{i0}^{t}$    & Capacity/speed allocated by IDC $i$ to DSJs in slot $t$ (Given vairiable) \\
 \ $\textbf{S}^t$  & Capacity/speed matrix in slot $t$ ($\textbf{S}^{\omega}$  for state $\omega$ ) \\
\ $r_{ij}$ &  Unit service rate by IDC $i$ for DTJ $j$    \\
\ $P_i^t$ & Power consumption of IDC $i$ in slot $t$  \\
\ $\alpha_i^t$ & Electricity price of IDC $i$ in slot $t$ ($\alpha_i^{\omega}$ for state  ${\omega}$)\\
\  $\phi_{ii^{'}}^t$   & Load shifting cost between IDC $i$ and $i^{'}$ in slot $t$  \\
\ $g^t()$   & Total cost function on $\textbf{S}^t$ in slot $t$ ($g^{\omega}()$ for state $\omega$) \\
\ $\pi_{\omega}$   & Distribution of system state $\omega$ (unknown to SSTF) \\
\  DTJ(DSJ)       & Delay tolerant (sensitive) jobs  \\
\hline
\end{tabular}\vspace{-5mm}
\end{center}
\end{table}

\subsection{Power consumption and cost model }
According to \cite{soda07}\cite{aotang}, power consumption of a
server (processor) running at a speed $s\in[0, 1]$  is
\begin{equation}
P(s)=\rho s^{\nu}+1-\rho, \label{powerserver1}
\end{equation}
where the  exponent $\nu\geq 1$, with a typical value of
2~\cite{aotang}, and $1-\rho$ represents  the power consumption in
the idle state, which is around $0.6$, and hardly lower than
0.5~\cite{gongchen}. In this paper, we choose  $\nu=2$, as in \cite{aotang}. Note that our schemes can be extended to the cases
with other values of  $\nu$.

Consider an IDC $i$. In a time slot $t$,  there are $K_i^t$ active
servers, and the total capacity demand is $S_i^t$. It can be shown that
the most energy-efficient  operation is to let each server evenly share
the demand, i.e., each server is running at a speed
$\frac{S_i^t}{K_i^t}$, which results in
 a total power consumption in slot $t$ of
\begin{equation}
P_i^t=(1-\rho)K_i^t+\frac{\rho {S_i^t}^2}{K_i^t},
\label{powerserver1}
\end{equation}
 where $S_i^t=S_{i0}^t+\sum_{j\in \Pi_i}S_{ij}^t$. Because we focus on trough-filling,  we take $K_i^t$  and $S_{i0}^t$ as given constants in each time slot.
We only control $S_{ij}^t$. Note that $P_i^t$ is a convex function of  $S_{ij}^t$.

Besides the power consumption of servers, other components in an IDC, e.g.,
 memory, I/O, hard disk, and non-IT equipments
  such as cooling systems, also contribute to
the total power consumption, which is roughly  proportional to that
by servers~\cite{apc}. Thus total power consumption of an IDC can be
obtained by scaling up $P_i^t$  with a constant factor.   For notation
brevity, we absorb this constant factor into the electricity price
at IDC $i$. Electricity price exhibits
significant diversity in both location and time. We use $\alpha_i^t$
to denote the price at IDC $i$  in time slot $t$. Although $\alpha_i^t$ is a time-varying variable, it varies slowly.
Typically, in a wholesale market, $\alpha_i^t$ is
determined  by Regional Transmission Organization (RTO) day-ahead based on expected load and changes hourly; or alternatively, $\alpha_i^t$ is
determined   in real-time (every 15min) based on the actual load.
 We consider energy cost of an IDC as the product of power consumption and its electricity price.
\subsection{Load shifting cost}
We also consider load shifting cost. In practice,  datacenter operators may have a lease with ISPs for
  data traffic among IDCs. Some large operators like
  Google
  and Microsoft may even have their own backbone networks to interconnect the IDCs. Either case, shifting cost is usually
  incurred during the acquisition or construction phase, which depends less on the traffic volume that the internal links carry~\cite{zheng}.
  However, since  DTJs have a lower priority, it is  desirable to schedule a limited link bandwidth to them. For example, when the time slot  is relatively long,  a higher utilization of the link capacity by DTJs will make the system more sensitive to the burst of DSJs, which enjoy  a higher priority on load shifting.  To prevent the increasing sensitiveness to DSJs, we use a piece-wise linear cost function
    with increasing rate to model the  shifting cost for DTJs.
 Let $\phi_{ii^{'}}^t$ denote the shifting cost in slot $t$ between
IDC $i$ and $i^{'}$, we have
 \begin{equation}
 \begin{split}
\phi_{ii^{'}}^t=\mbox{max}\left\{ a_{ii^{'}}^{\vartheta}\frac{\sum_{j\in\Upsilon_{ii^{'}}}D_{jii^{'}}^t}{B_{ii^{'}}^t}+b_{ii^{'}}^{\vartheta}  \right\},
\vartheta=\{1, 2,\ldots,\theta\},
 \label{powerserver1}
 \end{split}
 \end{equation}
where $\frac{\sum_{j\in\Upsilon_{ii^{'}}}D_{jii^{'}}^t}{B_{ii^{'}}^t}$ is the link capacity occupation ratio by DTJs. We have  $a_{ii^{'}}^{1}\leq\ldots a_{ii^{'}}^{\vartheta}\ldots\leq
a_{ii^{'}}^{\theta}$, which captures the increasing sensitiveness to capacity occupation ratio by DTJs.
$\phi_{ii^{'}}^t$ is a convex function on $D_{jii^{'}}^t$, and thus on $\textbf{S}^t$,   since it is the pointwise
maximum of a set of affine functions, and $D_{jii^{'}}^t$ is linear on  $\textbf{S}^t$. The model is also widely considered by previous works, e.g., in \cite{Fortz}. Note that our work can also
incorporate other shifting cost models with minor modifications.

\section{A benchmark scheme}
In this section, we first consider a benchmark scheme, where
the goal is to minimize the time average  of the
total cost of $N$ IDCs, including energy cost and shifting cost,  while stabilizing
the $M$  DTJ  queues. We name it stability-assured cost optimal trough-filling~(SCOTF).
In each time slot,  both the energy cost and the shifting cost are  functions of
$\textbf{S}^{t}$. The overall  cost in each slot also depends on $K_i^t$, $\alpha_i^t$, and $S_{i0}^t$, $i=1,\ldots, N$. Thus the overall cost is a time-varying function on $\textbf{S}^{t}$, denoted by $g^t(\textbf{S}^{t})$. Besides, capacity allocation and shifting constraints, i.e., $C_i^t$ and $B_{ii^{'}}^t$, are also time-varying. Thus
$\textbf{S}^{t}$ takes values in a time-varying set. Let $\Lambda^t$ denote the set of $\textbf{S}^{t}$ that satisfies capacity allocation and shifting constraints in slot $t$. SCOTF  is formulated as
\begin{eqnarray}
&\min_{\textbf{S}^{t}} \  \lim\mbox{inf}_{T\rightarrow \infty}\frac{1}{T}\sum_{t=1}^{T}g^t(\textbf{S}^{t})
 \notag \\
&\mbox{s. t.}\   \lim\mbox{sup}_{T\rightarrow \infty }  \frac{1}{T}\sum\nolimits_{t=1}^{T} Q_j(t)\leq \infty, \\
                   &\ \ \ \ \ \    \textbf{S}^{t}\in\Lambda^t, \ \   j=1,\ldots, M,
\label{general}
\end{eqnarray}
where the first constraint is to guarantee each DTJ queue's stability.
Note we use `sup' ( `inf' ) to guarantee the infinity exists.

It is difficult to solve problem (\ref{general}) directly in practice, because it is hard to obtain
prior system information of all time slots.
We present the problem of SCOTF here as a cost benchmark.  Our proposed schemes, one stochastic subgradient-based and one queue-based, require little system statistical information, and thus are more practical. The objectives of proposed schemes are not limited to
guaranteeing DTJ queue stability as in SCOTF. Good delay performance is also desired, especially for the queue-based scheme.

\section{Stochastic subgradient based trough filling}
\label{sec:ss}
We first consider an
ergodic scenario where system state  has a steady state distribution. Here a state characterizes
a unique set of all variables involved in the system, including $K_i^t$, $\alpha_i^t$, $S_{i0}^t$,  and $B_{ii^{'}}^{t}$, $i, i^{'}\in{1,\ldots, N}$. Let $\Omega$ denote the set of system states, and $\omega$  a generic system state, $\omega\in \Omega$,  $\pi_{\omega}$ the steady distribution of $\omega$, $g^\omega()$ is the cost function in state   $\omega$.
Let $\textbf{S}^{\omega}$ denote the capacity allocation matrix in state $\omega$, which is in the set $\Lambda^{\omega}$.
Let $\vec{\lambda}$ denote the mean of arrival rate vector of DTJs. SCOTF can be rewritten as
\begin{eqnarray}
&\mbox{min} \  g_{e}=\sum_{\omega\in \Omega}\pi_{\omega}g^{\omega}(\textbf{S}^{\omega})
 \notag \\
&\mbox{s. t.}\      \sum_{\omega\in \Omega}\pi_{\omega}\vec{R}^{\omega}({\textbf{S}^{\omega}})\geq \vec{\lambda} \notag \\
& \  \textbf{S}^{\omega}\in \Lambda^{\omega}
\label{ergodic},
\end{eqnarray}
We use $g_{e}^*$ to denote the optimal solution to the above problem, i.e., optimal cost in the  ergodic system
case, with the arrival rate $\vec{\lambda}$ stabilized. In practice,   $\vec{\lambda}$ can possibly be estimated by historic database or
prediction schemes. If the steady state distribution $\pi_{\omega}$ is available, then  (\ref{ergodic}) is a deterministic convex optimization problem.
However, in practice it may be difficult to obtain such  statistical knowledge.  We thus design a stochastic subgradient-based algorithm that can solve (\ref{ergodic}), without   prior information on $\pi_{\omega}$. Note the scheme  needs
the information of the average rate, i.e.,  $\vec{\lambda}$, or at least an upper bound to guarantee stability.

We first define a Lagrangian function associated with problem (\ref{ergodic})
 as
 \begin{equation}
 L(\vec{\mu},\vec{S})=\sum_{\omega\in\Omega}\pi_{\omega}g^{\omega}(\textbf{S}^{\omega})-\sum_{j=1}^{M}\mu_j(\sum_{\omega\in\Omega}\pi_{\omega}\sum_{i\in \Gamma_j }  r_{ij}S_{ij}^{\omega} -\lambda_j),
 \label{e1}
\end{equation}
where  $\vec{S}=\{\textbf{S}^{\omega}|{\omega\in\Omega}\}$,  $\textbf{S}^{\omega}\in \Lambda^{\omega}$, and $\vec{\mu}=(\mu_1,\ldots, \mu_M)$ is the set of the Lagrangian multipliers. Note $\vec{\mu}\geq 0$.
The dual problem of (\ref{ergodic}) is defined as
 \begin{equation}
\max_{\vec{\mu}>0} F(\vec{\mu}),
\label{dual}
\end{equation}
where
 \begin{equation}
F(\vec{\mu})=\min_{\vec{S}}L(\vec{\mu}, \vec{S}).
\end{equation}

To solve the dual problem, we first consider  (\ref{e1}). For a given multiplier $\vec{\mu}$, the problem is separable for
different states. Thus, we can solve the following problem  for a given state $\omega$,
 \begin{equation}
 \begin{split}
&\min_{\textbf{S}^{\omega}} g^{\omega}(\textbf{S}^{\omega})-\sum_{j=1}^{M}\mu_j(\sum_{i\in \Gamma_j }r_{ij}S_{ij}^{\omega} -\lambda_j) \\
& \mbox{s.t.}\ \ \  \textbf{S}^{\omega}\in \Lambda^{\omega}.
 \label{ep}
 \end{split}
\end{equation}
An examination  of (\ref{ep})  yields   the following optimization problem
of joint capacity allocation and load shifting after observing  system state in the current slot
\begin{eqnarray}
&\min\limits_{\textbf{S}^{\omega}}
\sum\limits_{i=1}^{N}\alpha_i^{\omega}\left[(1-\rho)K_i^{\omega}+\frac{\rho {(S_{i0}^{\omega}+\sum_{j\in \Pi_i}S_{ij}^{\omega})}^2}{K_i^{\omega}}\right]
+  \notag \\ & \sum\limits_{i=1}^{N}\sum\limits_{i^{'}\neq
i}\max_{1\leq\vartheta\leq\theta}\left\{ a_{ii^{'}}^{\vartheta}\frac{\sum_{j\in\Upsilon_{ii^{'}}}r_{i^{'}j}S_{i^{'}j}^{\omega}}{B_{ii^{'}}^{\omega}}+b_{ii^{'}}^{\vartheta}  \right\}  \notag \\&-\sum_{j=1}^{M}\mu_j( \sum_{i\in \Gamma_j }  r_{ij}S_{ij}^{\omega} -\lambda_j  )
\notag \\
&\mbox{s. t.}\ \ \  \sum\limits_{j\in\Pi_i}S_{ij}^{\omega} \leq  K_i^{\omega}-S_{i0}^{\omega}, i=1,\ldots,
N   \notag  \ \\ &
  \ \ \     \sum\limits_{j\in\Upsilon_{ii^{'}}}r_{i^{'}j}S_{i^{'}j}^{\omega} 
   \leq B_{ii^{'}}^{\omega}, i, i^{'}=1,\ldots, N, i\neq i^{'}. \label{ep3}
\end{eqnarray}
In (\ref{ep3}), the first item is the total energy cost, the second is the shifting cost, the first constraint is the capacity constraint on DTJs in IDC $i$ and the second constraint is bandwidth constraint between IDCs $i$ and $i^{'}$.
Clearly, (\ref{ep3}) is a convex optimization problem of $\textbf{S}^{\omega}$. This is because, the objective function
is the sum  of a set of convex and affine functions of $\textbf{S}^{\omega}$, and the constraints are both affine and thus
convex. We can solve it efficiently for
a given state $\omega$ in each time slot. When capacity
allocation is determined, load shifting policy is also jointly
determined, i.e., shift an amount of $r_{i^{'}j}S_{i^{'}j}^{\omega}$ for
DTJ $j$ from IDC $i$ to $i^{'}$ if $j\in \Upsilon_{ii^{'}}$.

     The dual problem can be solved using a stochastic subgradient algorithm~\cite{boyd}, which has the following iterative steps
\begin{equation}
\mu_j^{n+1}=[\mu_{i}+\beta^{n}\sigma_{j}^{n} ]^{+},
\label{iteration}
\end{equation}
where $n$ denote the $n$th iteration, i.e., $n$th time slots in our case, and $\vec{\sigma}^{n}=( {\sigma}_{1}^{n}, \ldots, {\sigma}_{M}^{n} )$ is the vector of stochastic subgradient that is chosen as
\begin{equation}
E( \vec{\sigma}^{n}|\vec{\mu}^0, \ldots, \vec{\mu}^n   )=\partial_{\vec{\mu}}F(\vec{\mu}^n),
\end{equation}
where $\partial_{\vec{\mu}}F(\vec{\mu}^n)$ is a subgradient of $F(\vec{\mu})$ at $\vec{\mu}^n$.
In this case, by updating  $\vec{\mu}^{n}$ using (\ref{iteration}), $\vec{\mu}^{n}$
converges to  the optimal solution of the dual problem (\ref{dual})  with probability 1, if the following conditions are satisfied
\begin{equation}
E( ({{\sigma}_{1}^{n}}^2+ \ldots+ {{\sigma}_{M}^{n}}^2)^{\frac{1}{2}}  |\vec{\mu}^0, \ldots, \vec{\mu}^n   )\leq c,
\label{condition}
\end{equation}
where $c$ is a constant, and $\sum_{n=0}^{\infty}{\beta^{n}}=\infty$, $\sum_{n=0}^{\infty}({\beta^{n}})^2=\infty$.
Note a candidate for $\beta^{n}$ can be $\frac{1}{n}$.

The subgradient   $\partial_{\vec{\mu}}F(\vec{\mu})$ can be a set, where by Danskin¡¯s Theorem~\cite{nonp}, we can choose a subgradient as
\begin{equation}
\partial_{\vec{\mu}_j}F(\vec{\mu})=-\sum_{\omega\in\Omega}\pi_{\omega}\sum_{i\in \Gamma_j }r_{ij}{S_{ij}^{\omega}}^{*} +\lambda_j, \ j=1,\ldots, M,
\label{gradient}
\end{equation}
where ${S_{ij}^{\omega}}^{*}$ is the optimal solution to problem (\ref{ep3}).
Note that  ${\sigma}_{j}^{n}$  is a stochastic subgradient if its expectation  equals to a subgradient. We can choose ${\sigma}_{j}^{n}$ as
\begin{equation}
 {\sigma}_{j}^{n}=-\sum_{i\in \Gamma_j }r_{ij}{S_{ij}^{\omega^n}}^{*} +\lambda_j, \ j=1,\ldots, M,
 \label{ss}
\end{equation}
where $\omega^n$ is the index of the system state at iteration $n$.
(\ref{condition}) is satisfied, because $r_{ij}{S_{ij}^{\omega^n}}^{*}$ is bounded, $\forall i, j$, which leads to bounded ${\sigma}_{j}^{n}$, $\forall j$.
 ${\sigma}_{j}^{n}$ defined in (\ref{ss})  is a  stochastic subgradient, because we consider an ergodic setting and thus the time average of ${\sigma}_{j}^{n}$  equals to the subgradient of (\ref{gradient}).
 Further, since the original problem (\ref{ergodic}) is a convex optimization problem  that satisfies the Slater's condition, there is no duality gap.

We name the above algorithm stochastic subgradient-based trough filling (SSTF).
SSTF converges to the optimal solution of problem (\ref{ergodic}). Thus it can achieve the optimal cost given a  service rate that assures queue stability. Note that SSTF can  work in non-ergodic settings. Lagrangian multiplier $\vec{\mu}$  has practical properties. It can be considered as  a price, which increases as service rate being smaller than the average arrival rate, i.e., capacity under-provisioning. In practice, by updating $\vec{\mu}$, SSTF can achieve good cost performance.
Moreover, the objective of SSTF is not limited to cost optimality only. One can tune the average service rate of SSTF, i.e., by adjusting $\vec{\lambda}$ in (\ref{ergodic}), to control the DTJ delay. Thus,
SSTF is NOT SCOTF in the ergodic setting.
 Another benefit of SSTF is that it also exploits temporal diversity of electrical prices.
 However, SSTF needs the knowledge of the average DTJ arrival rate, which may not be available in practice. Further, it
may converge slowly and it is difficult to characterize its delay performance.
 This motivates us to consider the following queue-based algorithm, which
 leverages queue information so that neither $\vec{\lambda}$  nor system distribution information is  required.

\section{Queue based trough filling}
\label{sec:qba}
\subsection{Algorithm Design}

In this section, we present  a  queue-based algorithm that explicitly considers queue backlog of DTJs.   The algorithm takes the instantaneous  system state (i.e., queue length, available server capacity and bandwidth, DSJ load demand) as the input. The algorithm also has a parameter to control the tradeoff between cost and queue delay.  We will also show that the algorithm  achieves
 bounded average queue backlog such that the system is stabilized, while the cost can be arbitrarily close to the optimal cost achieved by (\ref{ergodic}).

In each time slot $t$,  observe current queue backlog $Q_j(t), j=1,\ldots, M$,  $\alpha_i^t$,  $S_{i0}^t$,  $C_{i}^t$, and $B_{ii^{'}}^t$, $i=1,\ldots, N$. Allocate the capacity at each IDC $i$ for each queue $j$ according to the following optimization scheme, named queue-based trough filling (QTF):
\begin{eqnarray}
&\min\limits_{\textbf{S}^t}   -\sum\limits_{j=1}^{M}Q_j(t)\sum\limits_{i\in
\Gamma_j}r_{ij}S_{ij}^t+  \notag \\ &
V\sum\limits_{i=1}^{N}\alpha_i^t\left[(1-\rho)K_i^t+\frac{\rho {(S_{i0}^t+\sum_{j\in \Pi_i}S_{ij}^t})^2}{K_i^t}\right]
+  \notag \\ & V\sum\limits_{i=1}^{N}\sum\limits_{i^{'}\neq
i}\max_{1\leq\vartheta\leq\theta}\left\{ a_{ii^{'}}^{\vartheta}\frac{\sum_{j\in\Upsilon_{ii^{'}}}r_{i^{'}j}S_{i^{'}j}^{t}}{B_{ii^{'}}^{t}}+b_{ii^{'}}^{\vartheta}  \right\}\\
&\mbox{s. t.}\ \ \  \sum\limits_{j\in\Pi_i}S_{ij}^t \leq K_i^t-S_{i0}^t, i=1,\ldots,
N    \notag  \ \\ &
  \ \ \    \sum\limits_{j\in\Upsilon_{ii^{'}}}r_{i^{'}j}S_{i^{'}j}^t
   \leq B_{ii^{'}}^t, i^{'}=1,\ldots, N  \notag \\
  &
   \sum_{i\in\Gamma_{j}}r_{ij}S_{ij}^t\leq Q_j(t),
    j\in\{1, \ldots, M\}   \label{general4}.
\label{general2}
\end{eqnarray}
Similar to (\ref{ep3}),   (\ref{general4}) is a
convex optimization problem.  Thus at the beginning of each slot, capacity
allocation $\textbf{S}^t$ can be determined efficiently.

 The intuition of QTF is clear. When queue length $\sum\limits_{j=1}^{M}Q_j(t)$ is high, QTF has incentive to
 allocate a larger capacity to reduce the queue length. When the cost is relatively large or queue length is small, QTF is driven to
 allocate less capacity  to reduce the cost.  The control variable $V$
 is to balance the queue length and cost. If $V$ is large, QTF will result in lower cost but longer average queue delay.

 To better illustrate the intuition of the algorithm, we
 further consider a special case, where there is only one IDC with $M$ delay tolerant queues. In the single IDC case,
we can simplify  notations  by removing subscript $i$.  The capacity vector becomes $\textbf{S}^t=\{S_1^t,\ldots, S_M^t
\}$. We have the following scheme for capacity allocation, named single-IDC queue-based trough filling  (SQTF)
 \begin{eqnarray}
&\min\limits_{\textbf{S}^t}\    -\sum_{j=1}^{M}Q_j(t)r_jS_j^t+\\&
V\alpha^t\left[(1-\rho)K^t+\frac{\rho {(S_{0}^t+\sum_{j=1}^{M}S_{j}^t})^2}{K^t}\right]     \label{general1}\\
&\mbox{s.t.}\ \ \  \sum_{j=1}^{M} S_j^t\leq K^t-S_0^t     \ \\
  \ \ \ \ \ \ \ &S_j^t\geq 0, j=1, \ldots, M.
\label{single}
\end{eqnarray}
We have the following solution on $\textbf{S}^t$.

\textbf{Observation 1:} SQTF allocates  $\textbf{S}^t$ as:
\emph{ in each time slot $t$, choose the queue with the maximum  $Q_j(t)r_j$, denote as $j^{'}$}, then
\begin{equation}
\begin{split}
&S_{j^{'}}^t = \left\{\begin{array}{ll}
K^t-S_0^t, \ \mbox{if} \ Q_{j}(t)r_j\geq2V\rho\alpha^t   \\
 \frac{Q_j(t)r_jK^t}{2V\rho\alpha^t}-S_0^t, \       \mbox{elsif}\  Q_j(t)r_j\geq\frac{2V\rho\alpha^tS_0^t}{K^t}   \\
0, \mbox{else}
\end{array} \right.,\\
&S_{j^{}}^t=0 \ \mbox{if}\ j\neq j^{'}.
\label{power}
\end{split}
\end{equation}

 In other words,  SQTF is a threshold-based policy,  which serves the longest queue and only when its queue length is above a certain threshold.

\subsection{Performance analysis}
In this subsection, we analyze the performance of the QTF algorithm in terms of the cost and average delay performance.
Our analysis is based on Lyapunov drift optimization~\cite{mneely}.

Define  $r_i=\max\{ r_{ij}| j\in\Pi_i \}$, i.e., maximum unit service rate for all DTJs in IDC $i$. Let  $D_j^m$ denote the upper bound of arrival traffic size of DTJ $j$ in each slot.
We have the following proposition.

\textbf{Proposition 1:}  \emph{Assuming traffic of DTJs is i.i.d in each slot with mean $\vec{\lambda}$,
 the QTF algorithm   stabilizes the system for a given parameter $V$. In addition, an upper bound on average  queue length is
}
 \begin{equation}\scriptsize
 \lim_{T\rightarrow\infty}\frac{1}{T}\sum_{t=1}^{T}\sum_{j=1}^{M}E(Q_j(t))\leq
\frac{\sum_{i\in \cup\Gamma_j, \forall j}r_{i}^2{K_i^{max}}^2+ \sum_{j}{D_j^m}^2+Vg_{e}^{*}(\epsilon)}{\epsilon}
 \end{equation}
\emph{Further, average cost achieved by QTF, which has a cost denoted as  $g_q^t(\textbf{S}^t)$
in each slot $t$, is upper bounded as}
 \begin{equation}
\lim_{T\rightarrow\infty}\frac{1}{T}\sum_{t=1}^TE[g_q^t(\textbf{S}^t)]\leq
Vg_e^{*}+\frac{\sum_{i\in \cup\Gamma_j, \forall j}r_{i}^2{K_i^{max}}^2+ \sum_{j}{D_j^m}^2}{V}
 \end{equation}
\emph{where $g_{e}^{*}$ is the optimal solution to problem (\ref{ergodic}), and $\epsilon$ is a positive value,
$g_{e}^{*}(\epsilon)$ is the optimal solution to (\ref{ergodic}) with $\vec{\lambda}$ replaced by
$\vec{\lambda}+\textbf{1}\epsilon$. }

\emph{Proof:}  In the Appendix.

\section{Discussions}

\label{sec:joint}
\subsection{Joint DSJ and DTJ design }

Although SSTF and QTF are both proposed for trough-filling, with some modifications, they can be used for joint DSJ and DTJ capacity provisioning.  First,
$S_{i0}^t$, for DSJs, will become a part of the decision variables, together with $S_{ij}^t$ for DTJs.
 An important issue is how to guarantee service requirements for DSJs.

For SSTF, we can simply introduce a QoS constraint for DSJs. For example, if the slot length is large, i.e., tens of seconds to minutes,  following \cite{caltech1}, we can estimate the mean of DSJ rate for IDC $i$ in the beginning of the current slot, denoted  by $\lambda_{i0}^t$. Note that it is possible for $\lambda_{i0}^t$ to incorporate traffic from other IDCs due to certain traffic shifting schemes. Let $r_{i0}$ denote unit service rate for DSJs in IDC $i$. Following \cite{raolei},   a delay constraint can be imposed, e.g.,  $\frac{1}{r_{i0}S_{i0}^t-\lambda_{i0}^t}\leq \delta$, which is a linear constraint on   $S_{i0}^t$, and thus can be easily incorporated to our convex optimization problem.  When the time slot length is small, such as hundreds of milliseconds, it is unlikely to estimate mean of DSJ traffic in the current slot. In this case, one may assume DSJ traffic follows certain distributions  based on past measurement.
One can define outage probability as
a QoS constraint. That is, the probability that the load of DSJ in IDC $i$, i.e., $D_{i0}^t$, exceeds capacity  $S_{i0}^t$. The DSJ QoS constraint can be expressed as $\mbox{Pr}(D_{i0}^t>S_{i0}^t)\leq \delta_i$. Based on the knowledge of traffic distribution, e.g., Gaussian or exponential distribution,  one can rewrite the constraint function as a convex function of $S_{i0}^t$. Since time time slot length is small, outage probability can be easily measured. Adjusting $S_{i0}^t$ is probably necessary to eliminate the discrepancy between the real distribution of $D_{i0}^t$ and the assumed one using  stochastic approximation schemes.

Similar approaches can be applied to extend QTF.  For example, one can use the outage probability as a DSJ QoS constraint. Let $\delta_i$ denote outage probability constraint.  To enforce it, we can  design a virtual outage queue. Let $I_i(\cdot)$ as an indicator function. We have $I_i(t)=1$ if there is outage in slot $t$, i.e., $D_{i0}^t>S_{i0}^t$, and $I_i(t)=0$ otherwise. We use $O_i(t)$ to denote the virtual outage queue backlog in slot $t$, which updates as $O_i(t+1)=\mbox{max}\left\{ O_i(t)-\delta_i, 0  \right\}+I_i(t)$. It can be shown that the virtual queue is stable if
$\lim_{T\rightarrow\infty}\frac{\sum_{t=1}^{T} I_i(t)}{T}\leq \delta_i$, i.e., outage probability constraint satisfied. Note that $\delta_i$ can be considered  as the service rate of the virtual queue. Using the virtual outage queue, we can modify QTF to
provide capacity provisioning for DSJs. It is our future work to further investigate the joint design of capacity provisioning and QoS assurance for DSJs, and trough-filling.

\subsection{Implementation issues and caveats}
 In our schemes, the decision-maker needs to gather the input  in the beginning of each slot. The messaging delay is about tens of  milliseconds  \cite{Stanojevic}, and   each IDC only has a few parameters sent to the decision-maker.      Each time slot can be from several seconds to some minutes. Thus the messaging overhead is negligible.
    Note that the decision overhead is also negligible since the convex optimization problems can be solved efficiently. Load shifting overhead, i.e., network delay, can be easily constrained  by controlling the bandwidth for DTJs.

 In this paper, we consider homogenous servers for simplicity. However, in an IDC, servers may be different  in terms of power consumption, maximum speed, and memory.  To apply our schemes, we can further classify the servers to different units. Homogenous or similar servers belong to one unit.  The input is no longer IDC-based, but unit-based.  In practice, we can simply   classify servers according to their ages. Typically, there are three stock-keeping units (SKUs) in an IDC, i.e., latest, one-year-old, and two-year-old.

In practice, some DTJs may need to be finished by a  deadline. Different classes of DTJs may have different deadlines.   Designing  energy-efficient DTJ scheduling algorithms with heterogenous deadlines for IDCs
is an interesting open problem. We will consider it in the future.

In this paper, we mainly focus on CPU-intensive DSJs. We will also extend our work to
I/O intensive DTJs. Besides, we will also explicitly consider the effect of virtualization, by which
   performance versus
power curve may become  more difficult to quantify~\cite{Shekhar}\cite{Stanojevic}.

\section{Performance evaluation}
\label{sec:evaluation}

In this section, we evaluate the performance of SSTF and QTF, using  both synthetic and real traces.

\subsection{ Synthetic traces based simulation}
\subsubsection{Simulation setup}
We consider five IDCs in different locations. There are totally ten DTJ queues randomly originated in one of
the five IDCs. The IDC set $\Gamma_j$ that can serve a DTJ $j$ is chosen randomly. Idle power consumption $1-\rho$
 is set as 0.5. To create an ergodic setting, we set 100 states, in each of which we set different total capacity, load shifting constraint, demand by DSJs, and electricity prices. Capacity of each IDC is uniformly distributed from 10k to 15k. Load shifting constraint is uniformly distributed from 3000 to 4000. Load shifting cost parameters are set the same as  in \cite{Fortz}.  Electricity price is uniformly distributed from 1 to 10.
DSJ demand, set as a ratio of the total capacity, is  randomly distributed from 0 to 0.4.  Thus average DSJ demand is about 20\% of the total capacity. We consider different ratios between the load of DTJ and DSJ, by setting different average arrival rates of DTJs. The ratios are 0.5, 1, 1.5, 2, 2.5, 3, and 3.5, respectively. Thus the percentage of  DTJ demand in  the total capacity ranges from 10\% to 70\%.   We simulate 100k time slots in each of the 30 simulation settings. In different time slots, a system state is chosen randomly according to a predefined probability.

\begin{figure}
\centering {
\includegraphics[width=3.5in, height=1.8in]{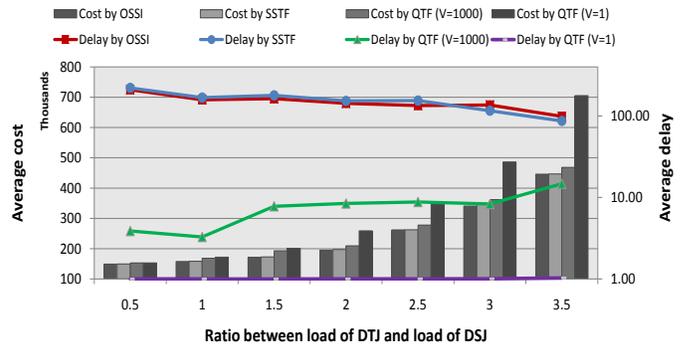}
} \vspace{-0.4cm}
\caption{Delay and cost of different schemes with different ratio between load of DTJ and DSJ. } \label{ratio}
\end{figure}
\subsubsection{Simulation results}
We first compute the Optimal Solution to (\ref{ergodic}) with the System distribution Information, which is difficult to obtain in practice. We name it OSSI and compare it with SSTF and QTF. First, by Fig.~\ref{ratio}, we observe that the cost of SSTF is very close to that of OSSI, under different DTJ load ratios. Their queue delays are also very close. In this paper, since we also consider idle power consumption, i.e., $(1-\rho)K_{i}^t$, and DSJ power consumption. When load of DTJs  is low, such as with the ratios of 0.5 and 1, costs of different schemes are very close because the impact of DTJs is small.
To study the convergence of SSTF, we also consider the DTJ power consumption separately. Results show that
SSFT and OSSI achieve very close performance in terms of cost and delay.  We do not plot results here due to the page limit.
In Fig.~\ref{ratio},   we consider QTF with $V=1$ and $V=1000$, respectively. For both cases, we see QTF leads to a higher cost, but the queue delay is significantly smaller compared to that by OSSI and SSTF. QTF with $V=1000$ has a slightly larger cost than   OSSI and SSTF, but much smaller delay, even when  DTJ load is high, e.g., with a ratio of 3.5. In this case,  QTF with $V=1$ has a very small delay, i.e., almost 1, with a much higher cost.  Thus, in practice, one can tune the value of $V$ to obtain a desirable tradeoff between cost and delay, especially when load of DTJs is high.
\begin{figure*}
\centering{
\includegraphics[width=2.3in, height=1.5in]{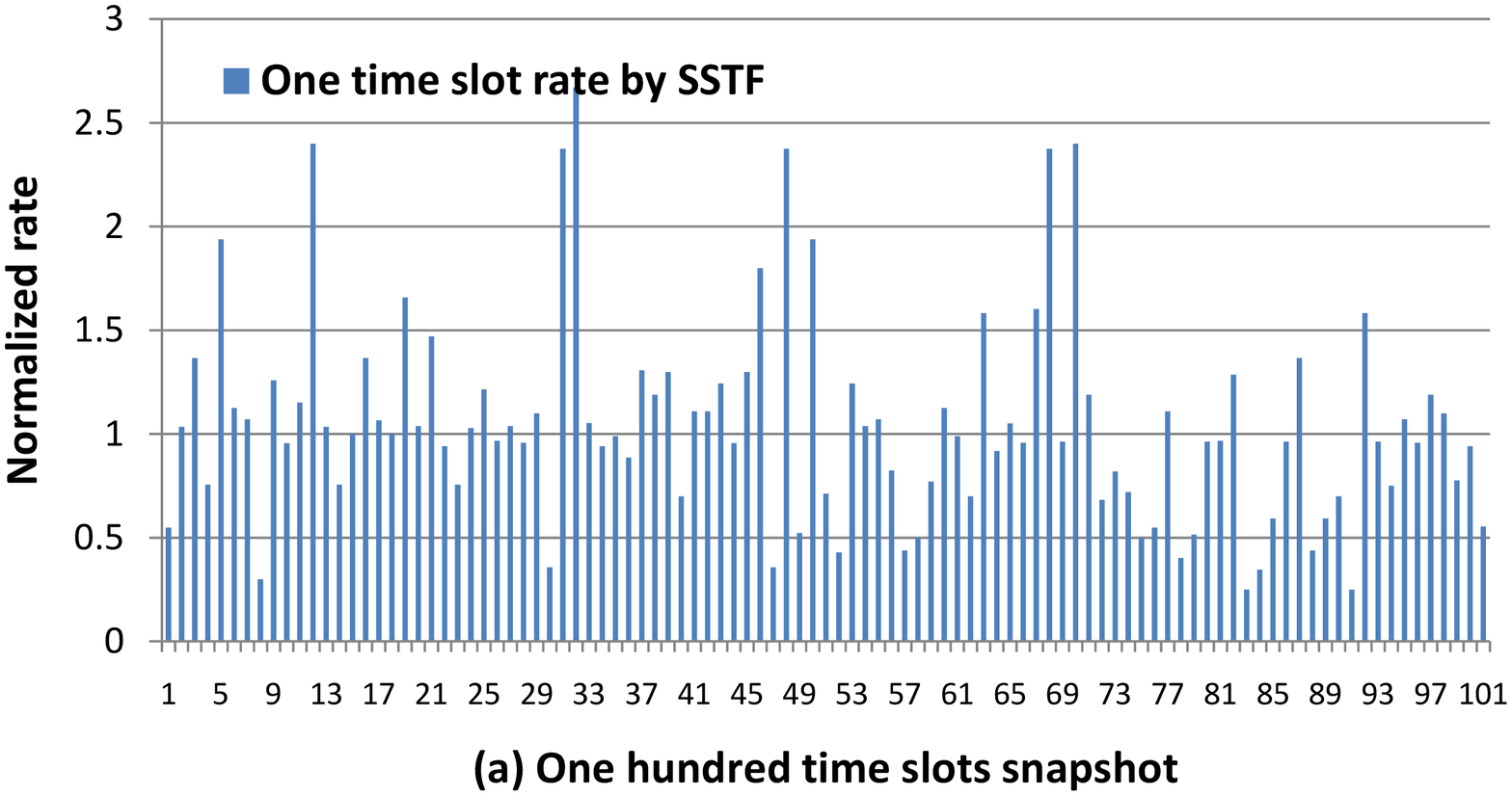} \label{rate1}
\includegraphics[width=2.3in, height=1.5in]{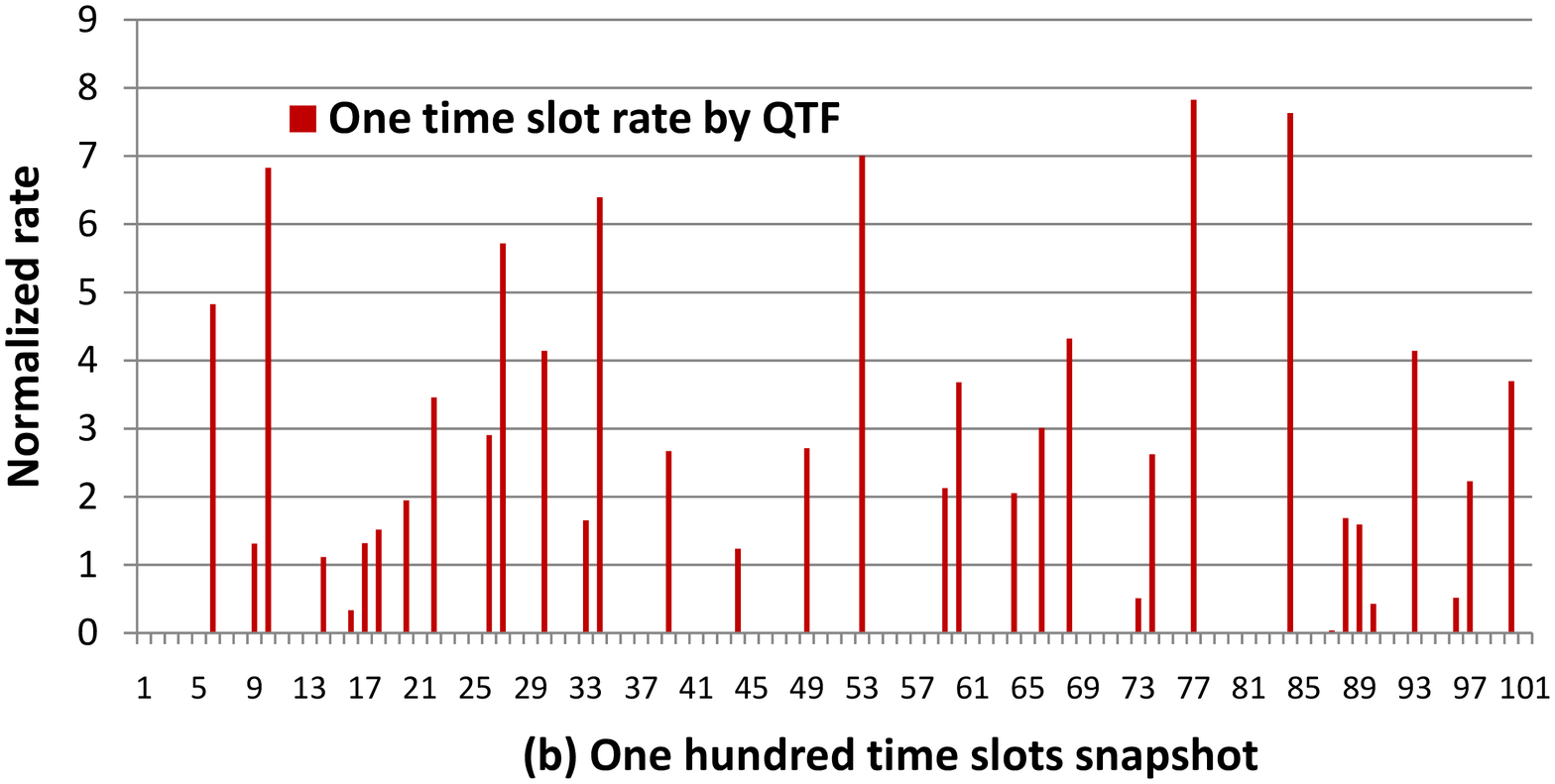}\label{rate2}
\includegraphics[width=2.3in, height=1.5in]{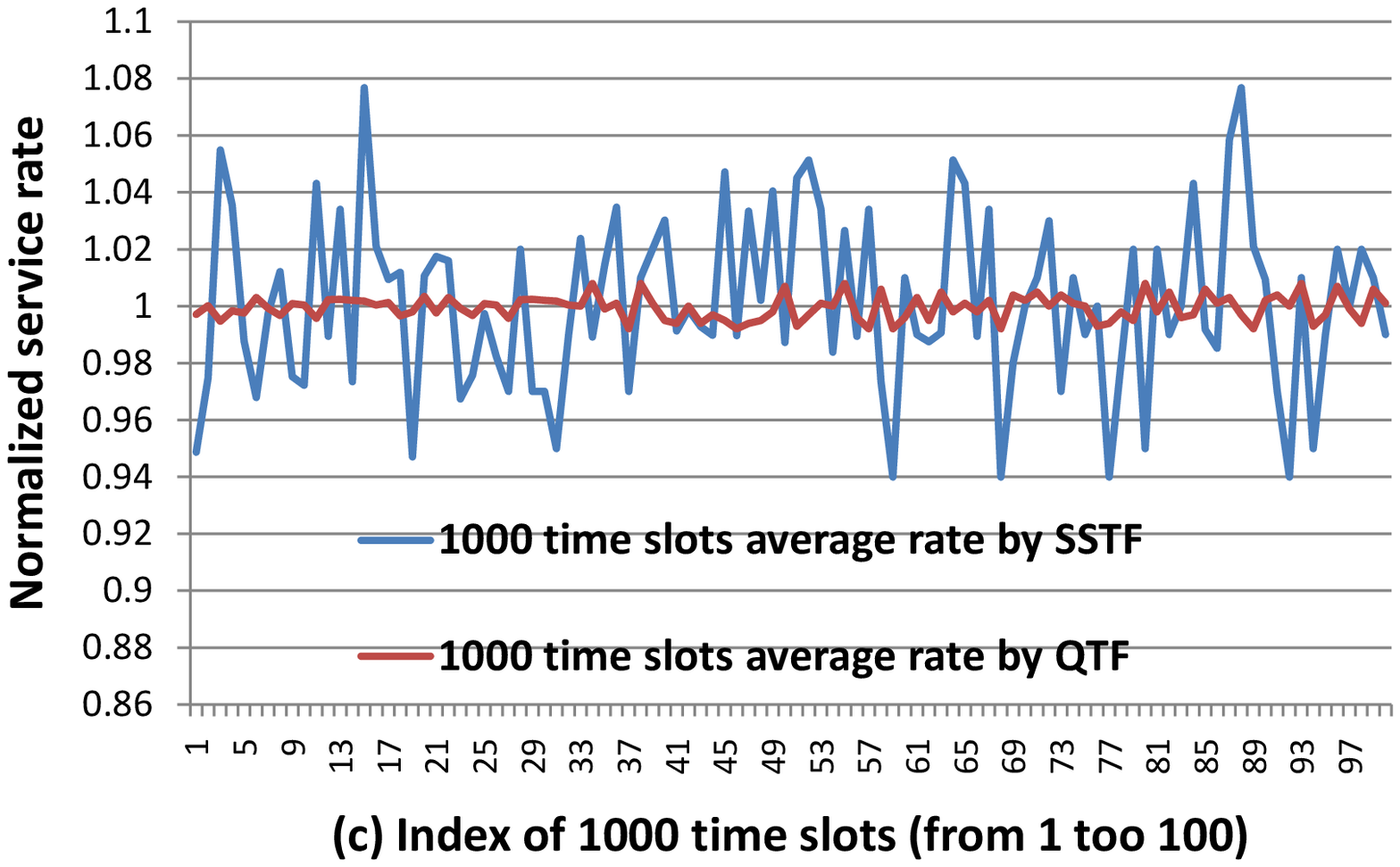}\label{rate3}
}
\caption{Rate assignment by SSTF and  QTF on different time resolutions.} \label{rate}
\end{figure*}

In Fig.~\ref{ratio},  the queue delay of OSSI and SSTF is very large, which  holds even when we    set the average service rate (slightly) larger than the  arrival rate.
We examine the service rates of a DTJ queue in different time resolutions to find the reasons. We first consider one slot service rate, normalized over the average DTJ arrival rate.  We plot 100 slots rate in Fig.~\ref{rate}{a} and (b), for SSTF and QTF, respectively (Service rate by OSSI is very similar to that by SSTF). Note in Fig.~\ref{rate}, the ratio of DTJ load is 1 and $V$ for QTF is 1000.  It is observed that rate assignment by SSTF is quite even for each slot. The DTJs always receive a service rate in each slot. Rate assignment by QTF is much more bursty. Service rate is non-zero only by every several slots. This result is consistent to Observation 1 where we show capacity allocation by QTF for a single IDC is a threshold-based policy based on the queue length.   In the time slots without being served, jobs accumulate and queue delay increases. This is the reason that there is a queue delay about 5 in Fig.~\ref{ratio} for QTF ($V=1000$ and DTJ load ratio of 1). Nevertheless, queue stability is guaranteed since service rates are fairly large every several slots such that jobs accumulated can be finished. We also examine a large time resolution rate, i.e., average rate over every 1000 time slots (normalized over average arrival rate). We plot results in Fig.~\ref{rate}(c). An interesting observation is that in this case, rate by SSTF is more bursty than that by QTF.  Then during the periods that normalized service rates are lower than 1, DTJs accumulate such that queue length is fairly large in most slots. Although jobs can be finished during periods when service rates are large than 1, significant  delay cannot be avoided.

One can increase average service rates of   SSTF to obtain a smaller delay. But much more capacity needs to be consumed, which results in much higher cost. In many cases when load of DTJ is high, there is little space for SSTF to increase service rates. QTF can lead to arbitrary delay by tuning $V$. One important property of QTF is that no matter $V$ is large or small, the average service rate of QTF is always close to arrival rate, because it leverages the queue information. Thus QTF provides a more efficient method in saving cost and reducing delay.   There are other findings, such as load shifting also plays an important role in reducing cost and queue delay. Due to the page limit, we omit them here.

\begin{figure}
\centering {
\includegraphics[width=3.5in, height=1.8in]{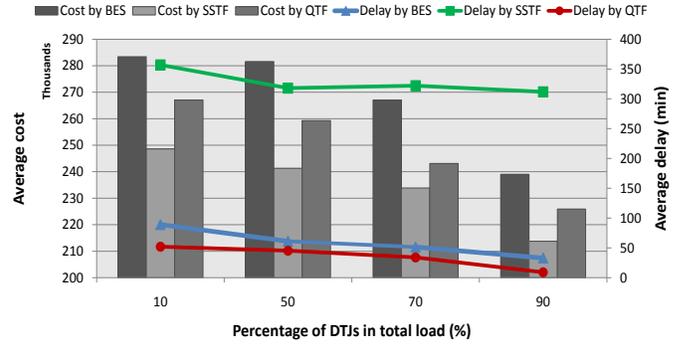}
} \vspace{-0.4cm}
\caption{Delay and cost by SSTF and QTF on real traffic trace } \label{trace}
\end{figure}

\subsection{ Real trace based simulation}

In this subsection, we use real datacenter traffic trace to study the performance of SSTF and QTF. Our trace comes from a  commercial datacenter operated by a large
cloud service provider in U.S. We obtain a  Hadoop distributed file system (HDFS) log  for one datacenter for thirty days.
 The HDFS log records the information of all received packets, including the packet size and time-stamp.
The original data does not differentiate DSJs and DTJs (In fact, to differentiate such traffic without application-layer information is itself a challenging issue in practical data center operations,  which is an active research topic  itself.). To address this issue,
 we simply adopt a threshold-based policy.
We assume that a large packet is likely to be delay tolerant, and treat a packet with a size larger than a certain threshold as DTJ.
 This classification is rational, as authors in \cite{VL2} indicate that
 most Internet request such as searching and web browsing are
 are only a few kb in size.
  We set threshold as 10, 50, 100, and 150Mb, to obtain different ratios between  DSJ load and DTJ load, which results in
   the percentage  of DTJ load in the total load roughly as  90\%, 70\%, 50\%, and 10\%, respectively. Note here we assume one unit (Mbit) of DSJs requires one unit of capacity, and  one unit of DTJs requires 0.133 unit of capacity on average,  by the same rate setting as the above simulations (average unit rate $r_{ij}$ is roughly 7.5).

 To simulate multiple IDCs and multiple DTJ queues, we choose twenty days of large
 packet traces as   ten DTJ traffic traces, so that each of them has  a two-day traffic trace. We choose ten days of small packet traces as the demand of DSJ for five IDCs considered.  We consider a time slot length as 20 seconds. Therefore we have 8640 time slots for each two-day traffic trace.

  Further, we use  the electricity data in five wholesale market regions in 02/22/2011. They are California~(Hub SP 15-EZ), Louisiana~(Entergy),
  New England~(NEPOOL Mass), Pennsylvania~(PJM West), and Texas~(ERCOT SOUTH).
  The capacity is  uniformly distributed between $1000$ and $1200$. The bandwidth constraint is uniformly distributed between 1000 and 1500. The other setting is the same as in the synthetic traffic case.

  We compare SSTF and QTF to the best effort service scheme (BES). In each slot, BES serves as much demand as possible for DTJ queue, in a best-effort fashion.
  When the available capacity in an IDC is not enough to finish current jobs, it equally shares the capacity among all DTJ queues. In the simulation, we assume SSTF knows the average DTJ arrival rate. The average service rate of SSTF is set equal to the  average DTJ arrival
  rate. The control variable of QTF is set to 1000. We observe from Fig.~\ref{trace} that for different percentages of DTJ load,
 BES always leads to the highest cost, while SSTF always has the lowest cost.     The  delay of SSTF is  large, almost 5 hours. One reason is that  it explores temporal electrical price diversity  in a large time scale.  One may think that BES would result in the lowest delay.  But in Fig.~\ref{trace}, average delay of BES is always larger than that of QTF. The reason is that load shifting is not used in BES. Thus  queues suffer large delay in an IDC with less available capacity. This illustrates that load shifting is not only necessary in reducing cost, but also important in exploring available capacity to improve delay performance. In summary, in Fig.~\ref{trace}, we observe that QTF is efficient in both saving cost and reducing delay.

 It is also observed that as the percentage of DTJ load increases, the total cost  decreases and the average DTJ delay also decreases. The reason is that when DSJ load decreases, total load amount decreases as DSJ requires more capacity per unit traffic. More capacity is thus available for DTJ, which leads to a smaller DTJ  delay and more space for energy saving.

\section{Conclusions}
\label{sec:con} In this paper,
we study intelligent trough filling that achieves both energy efficiency and good delay performance.
We design joint dynamic speed scaling  and load shifting schemes.
We first present   a stochastic subgradient based trough filling algorithm, named SSTF, which solves a convex optimization
problem for capacity allocation and load shifting in each slot.  SSTF does not need the information of underlying distribution of system state.
The SSTF can converge to optimal cost with a certain service rate constraint.
We further propose a queue-based trough filling  algorithm, named QTF, which also solves a convex optimization
problem for capacity allocation and load shifting in each slot. We show QTF can achieve optimal tradeoff between queue delay and  cost. Our extensive simulations based on both synthetic and real datacenter traces show that SSTF achieves optimal cost, but has a large delay.
 QTF achieves both desirable cost and delay. In practice, SSTF can be applied to the scenario where DTJs can have a large time delay, e.g., half of a day.   QTF can be applied to the case where smaller time delay is desirable, e.g., tens of minutes.

\appendix
To prove Proposition 1, we first need Lemma 1 as

\textbf{Lemma 1:}   \emph{For the optimization problem
(\ref{ergodic}), with $\vec{\lambda}$ replaced by
$\vec{\lambda}+\textbf{1}\epsilon$, the resulting optimal solution
$g_{e}^{*}(\epsilon)$ reaches $g_{e}^{*}$ as $\epsilon$ reaches 0.
}

\emph{Proof:}
 We write the Lagrangian of problem
(\ref{ergodic}) as
 \begin{equation}
 L(\vec{\mu},\vec{S})=\sum_{\omega\in\Omega}\pi_{\omega}g^{\omega}(\textbf{S}^{\omega})-\sum_{j=1}^{M}\mu_j(\sum_{\omega\in\Omega}\pi_{\omega}\sum_{i\in \Gamma_j }  r_{ij}S_{ij}^{\omega} -\lambda_j)
 \label{ea}
\end{equation}
When $\vec{\lambda}$ replaced by
$\vec{\lambda}+\textbf{1}\epsilon$, we have $L(\vec{S}, \lambda+\textbf{1}\epsilon, \vec{\mu})\rightarrow
L(\vec{S}, \lambda, \vec{\mu})$ as $\epsilon\rightarrow 0$. Since (\ref{ergodic})
is a convex optimization problem. We have $g_{e}^{*}(\epsilon)$ reaches $g_{e}^{*}$ as $\epsilon$ reaches 0.

We next present proof to Proposition 1.

\emph{Proof:}
Consider the $M$ DTJ queues $\vec{Q}(t)=(Q_1(t),
\ldots,Q_M(t))$. We introduce a non-negative Lyapunov function as
$L(\vec{Q}(t))=\sum_{j=1}^{M}Q_j^2(t)$. Define one-slot Lyapunov
drift as
\begin{equation}
\Delta(t)=E\left\{L(\vec{Q}(t+1))-L(\vec{Q}(t))| \vec{Q}(t)
\right\}
 \label{d1}
\end{equation}
In terms of the fact that $(\mbox{max}[a-b, 0]+c)^2\leq
a^2+b^2+c^2+2a(c-b)$, for any $a, b, c\geq 0$, we have
\begin{equation}
\begin{split}
&Q_j^2(t+1)- Q_j^2(t)\leq \\ &\sum_{i\in \Gamma_j}(r_{ij}S_{ij}^t)^2+{D_{j}^t}^2+2Q_j(t)(D_{j}^t-\sum_{i\in \Gamma_j}r_{ij}S_{ij}^t), \forall j
\label{d2}
\end{split}
\end{equation}
Based on (\ref{d2}), we further have
\begin{equation}
\begin{split}
\Delta(t)\leq
&E\left[\sum_{j=1}^{M}\sum_{i\in\Gamma_j}{(r_{ij}S_{ij}^t)^2}|\vec{Q}(t)\right]+E\left[\sum_{j=1}^{M}{D_j^t}^2|\vec{Q}(t)\right]+
\\& 2E\left[\sum_{j=1}^{M}Q_j(t)\left(D_j^t-\sum_{i\in \Gamma_j}r_{ij}S_{ij}^t
\right)|\vec{Q}(t) \right]. \label{d30}
\end{split}
\end{equation}
Note  $\sum_{j=1}^{M}\sum_{i\in\Gamma_j}{r_{ij}^2{S_{ij}^t}^2}$
is bounded  by $\sum_{i\in \cup\Gamma_j, \forall j}r_{i}^2{K_i^{max}}^2$, where $r_i=\max\{ r_{ij}| j\in\Pi_i \}$, i.e, the maximum service rate with full server capacity.
 In each slot, we also have assumed that the arrival traffic size of each DTJ $j$
is bounded by $D_j^m$.

For brevity, here we define $B=
\sum_{i\in \cup\Gamma_j, \forall j}r_{i}^2K_i^2+\sum_{j}{D_j^m}^2$.
Since traffic of DTJs in each slot is independent of
queue backlog $\vec{Q}(t)$,  we
can rewrite (30) as
\begin{equation}\small
\begin{split}
\Delta(t)\leq B+ 2\sum_{j=1}^MQ_j(t)\lambda_j
-2E\left[\sum_{j=1}^{M}Q_j(t)\sum_{i\in \Gamma_j}r_{ij}S_{ij}^t
|\vec{Q}(t) \right]. \label{d4}
\end{split}
\end{equation}

We consider the drift-plus-cost for the system where cost is
resulted by QTF.
 The cost is the
expected cost that is conditional on queue backlog in time slot $t$,
which  can be written as $E(g_q^t(\textbf{S}^t)|\vec{Q}(t))$.
Note $V$ is a control variable, we have
\begin{equation}
\begin{split}
&\Delta(t)+VE[g_q^t(\textbf{S}^t)|\vec{Q}(t) ]\leq B+
2\sum_{j=1}^MQ_j(t)\lambda_j\\&
-2E\left[\sum_{j=1}^{M}Q_j(t)\sum_{i\in
\Gamma_j}r_{ij}S_{ij}^t |\vec{Q}(t) \right]+VE[g_q^t(\textbf{S}^t)|\vec{Q}(t) ].\\
\label{d5}
\end{split}
\end{equation}
By (\ref{d5}), we can see that QTF minimizes drift-plus-cost in each time slot.
Thus we have
\begin{equation}\small
\begin{split}
&2\sum_{j=1}^MQ_j(t)\lambda_j+2E\left[Vg_q^t(\textbf{S}^t)-\sum_{j=1}^{M}Q_j(t)\sum_{i\in
\Gamma_j}r_{ij}S_{ij}^t |\vec{Q}(t) \right]\\
&\leq 2\sum_{j=1}^MQ_j(t)\lambda_j
+2E\left[Vg_{e}^{*}(\epsilon)-\sum_{j=1}^{M}Q_j(t)(\lambda_j+\epsilon)
|\vec{Q}(t) \right]
\\=&-2\epsilon\sum_{j=1}^{M}Q_j(t)+Vg_{e}^{*}(\epsilon).
\label{d6}
\end{split}
\end{equation}

By (\ref{d5})(\ref{d6}), we have
\begin{equation}
\begin{split}
&\Delta(t)\leq
B-2\epsilon\sum_{j=1}^{M}Q_j(t)+Vg_{e}^{*}(\epsilon)-VE[g_q^t(\textbf{S}^t)|\vec{Q}(t)]. \label{d7}
\end{split}
\end{equation}
Taking expectations of drift $\Delta(t)$ with respect to the distribution of the random queue backlog
$\vec{Q}(t)$ at time $t$, we have
  \begin{equation}
\begin{split}
&E\left[L(\vec{Q}(t+1))- L(\vec{Q}(t)) \right] \leq \\ &
B-2\epsilon\sum_{j=1}^{M}E[Q_j(t)]+Vg_{e}^{*}(\epsilon)
-VE[g_q^t(\textbf{S}^t)].
\label{d8}
\end{split}
\end{equation}

The above inequity is satisfied for all time slot $t$. Summing the
$\Delta(t)$ over time slot $t=1, 2,\ldots, T$, we have
\begin{equation}
\begin{split}
&E\left[L(\vec{Q}(T))- L(\vec{Q}(1)) \right] \leq \\ &
TB-2\epsilon\sum_{t=1}^{T}\sum_{j=1}^{M}E[Q_j(t)]+TVg_{e}^{*}(\epsilon)
-V\sum_{t=1}^{T}E[g_q^t(\textbf{S}^t)].
\label{d9}
\end{split}
\end{equation}

By (\ref{d9}), we can get
\begin{equation}
\begin{split}
\frac{1}{T}\sum_{t=1}^{T}\sum_{j=1}^{M}E(Q_j(t))\leq
\frac{B+Vg^{*}(\epsilon)}{\epsilon}+\frac{L(\vec{Q}(1))}{T\epsilon}.
\end{split}
\end{equation}
As $T\rightarrow\infty$, we have
$\lim_{T\rightarrow\infty}\frac{1}{T}\sum_{t=1}^{T}\sum_{j=1}^{M}E(Q_j(t))\leq
\frac{B+Vg_{e}^{*}(\epsilon)}{\epsilon}$. Thus the queue backlog is
bounded and system stability holds. Further
\begin{equation}
\begin{split}
\frac{1}{T}\sum_{t=1}^{T}E[g_{q}^{t}(\textbf{S}^t)]\leq
g_{e}^{*}(\epsilon)+\frac{B}{V}+\frac{L(\vec{Q}(1))}{TV}.
\label{relation}
\end{split}
\end{equation}
As $T\rightarrow\infty$, we have
$\lim_{T\rightarrow\infty}\frac{1}{T}\sum_{t=1}^{T}E[g_q^t(\textbf{S}^t)]\leq
g_{e}^{*}(\epsilon)+\frac{B}{V}$.
Since by Lemma 1, we have $g_{e}^{*}(\epsilon)\rightarrow g_{e}^{*}$
as $\epsilon$ reaches 0.  (\ref{relation}) is independent of
$\epsilon$. Thus we have $\lim_{T\rightarrow\infty}\frac{1}{T}\sum_{t=1}^{T}E[g_q^t(\textbf{S}^t)]\leq
g_{e}^{*}+\frac{B}{V}$ holds.

\end{document}